\documentclass[12pt]{article}

\usepackage[left=2.5cm,top=3.5cm,right=2.5cm,bottom=2cm,nohead]{geometry}

\usepackage{setspace}

\usepackage{url}
\usepackage{multirow}
\usepackage{multicol}
\usepackage{graphicx}
\usepackage{caption}
\usepackage{subcaption}
\usepackage[export]{adjustbox}
\usepackage{color}

\newcommand{\aspas}[1]{{``#1''}}

\usepackage{booktabs}
\usepackage{framed}
\usepackage{tcolorbox}

\begin{document}

\title{How do Developers Promote Open Source Projects?}

\author{Hudson Borges, Marco Tulio Valente \\
	{\normalsize Department of Computer Science, UFMG, Brazil} \\
	{\small \{hsborges,mtov\}@dcc.ufmg.br}
}

\date{}

\maketitle

\begin{abstract}
\noindent Open source projects have an increasing importance on modern software development. For this reason, these projects, as usual with commercial software projects, should make use of promotion channels to communicate and establish contact with users and contributors. In this article, we study the channels used to promote a set of 100 popular GitHub projects. First, we reveal that Twitter, user meetings, and blogs are the most common promotion channels used by the studied projects. Second, we  report a major difference between the studied projects and a random sample of projects, regarding the use of the investigated promotion channels. Third, we show the importance of a popular news aggregation site (Hacker News) on the promotion of open source. We conclude by presenting a set of practical recommendation to open source project managers and leaders, regarding the promotion of their projects.
\end{abstract}

\section{Introduction}
\label{sec:intro}

Open source projects have an increasing importance in modern software development.
For example, several open source projects are daily used by millions of users.
However, it is very important to continually attract more participants and contributors to these projects, in order to increase the chances of long-term success~\cite{comino2007}.
Particularly, several channels can be used to promote open source software, helping to keep the interest of the community and also to attract new members.

In this article, we investigate the most common channels used by developers to promote open source projects.
We manually inspected a large set of popular projects on GitHub, which is the world's largest collection of open source software, with around 27 million users and 77 million repositories~\cite{githubsearch}.
Our contributions include: (i) data about the promotion channels frequently used by popular open source projects; (ii) a comparison on the use of promotion channels by popular projects and by random ones; and (iii) an analysis of the impact of promotion on Hacker News, a popular news aggregation site, in the popularity of the studied projects.
Our findings help practitioners to understand the importance of using promotion channels in the open source development context.

\section{Study Design}
\label{sec:design}

To reveal the most common promotion channels used by developers, we manually inspected the documentation of the top-100 projects with most stars on GitHub (stars is a popular feature to manifest interest or satisfaction with GitHub projects~\cite{icsme2016}).
We restricted our analysis to popular projects because they have a large number of users and therefore need better and efficient ways to communicate with users and also to attract new contributors.

\begin{figure}[!ht]
	\center
	\includegraphics[width=0.65\textwidth,keepaspectratio,trim={0 2em 0 2em},clip]{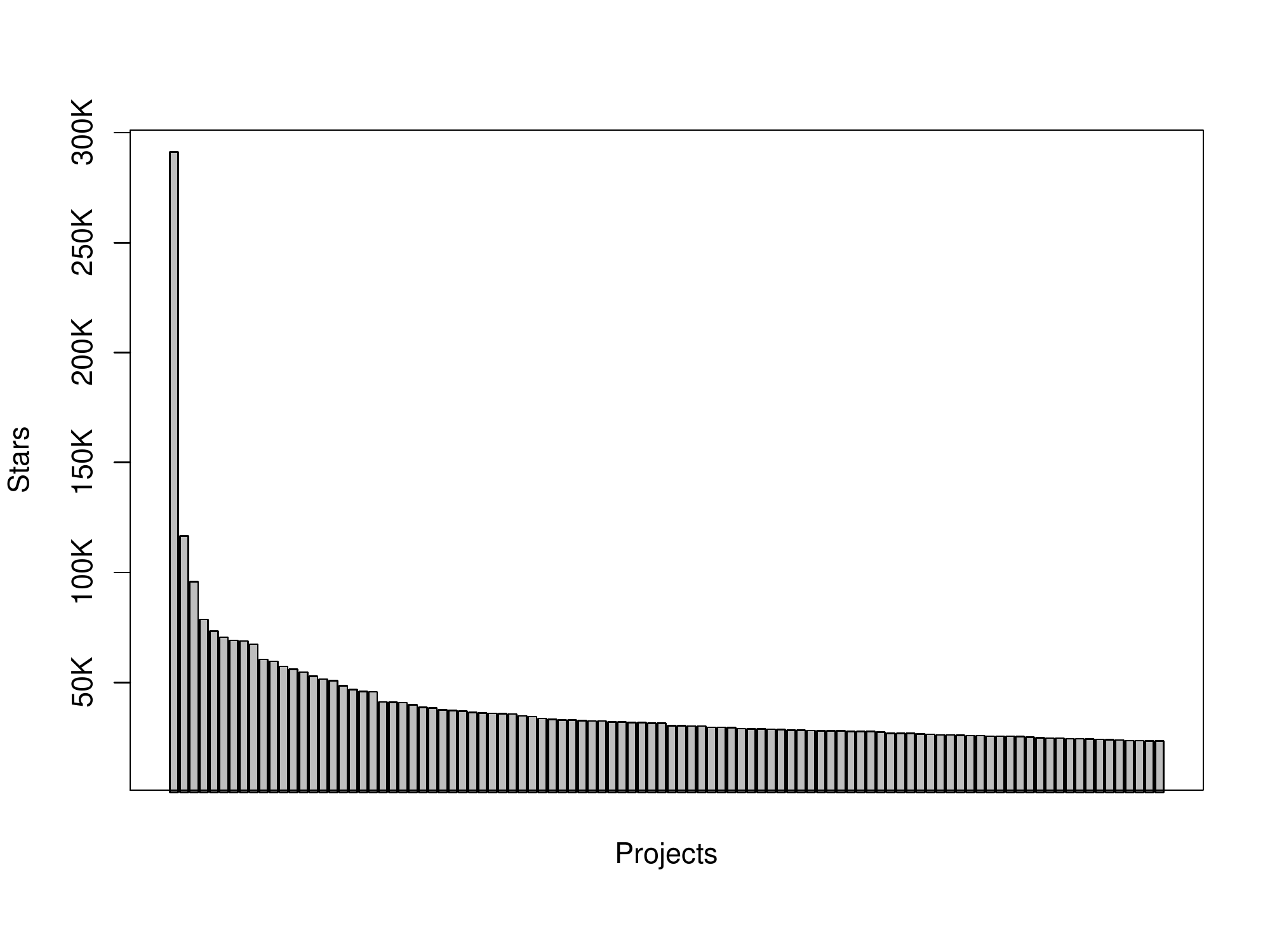}
	\caption{Number of GitHub stars of the analyzed projects}
	\label{fig:repos-overview}
\end{figure}

Figure~\ref{fig:repos-overview} shows the distribution of the number of stars of the projects considered in this study.
This number ranges from 291,138 stars ({\sc \mbox{freeCodeCamp/freeCodeCamp}}) to 23,322 stars ({\sc \mbox{tiimgreen/github-cheat-sheet}}).
The considered projects are primarily developed on 17 programming languages; JavaScript is the most common one (40 projects), followed by Python (9 projects) and Go (5 projects).
Furthermore, 14 projects only include markdown files with documentation purposes (e.g., projects with tutorials, books, awesome lists, etc).
Finally, regarding the project owners, 69 are organizational accounts and 31 are user accounts.\medskip

For each of these 100 projects, the first author of this paper initially inspected their READMEs on GitHub to identify the channels used to promote the projects and to keep the users up-to-date with important information about them.
For example, the following sentence is available on the README of {\sc adobe/brackets}: \aspas{\it You can see some screenshots of Brackets on the \underline{wiki}, intro videos on \underline{YouTube}, and news on the Brackets \underline{blog}}.
In this case, wiki and YouTube are used to support users whereas blog is a channel used to disseminate news about {\sc Brackets}.
Thus, only blog is considered a promotion channel in our study.
Next, we inspected the projects' website, for those projects having one.
We navigated through the site pages, searching for more channels used to promote the projects.

After this manual inspection, the following promotion channels emerged:

\begin{itemize}
	\item {\bf Blogs}, which are used, for example, to publish announcements of new software versions, upcoming events, and improvements.

	\item {\bf Events and Users Meetings:} Organizing events and supporting users meetings are other strategies commonly followed to promote projects. On events the initiative usually comes from the development team or from the organization that supports the project, whereas on user meeting the initiative comes from the users, usually from a specific region or country. We rely on Meetup (\url{https://meetup.com}) to discover users meetings.

	\item {\bf Twitter, Facebook, and Google+}, which are also used to connect the projects to users. We considered only official accounts, which are explicitly advertised on the project documentation or are verified by the social network (e.g., \url{https://support.twitter.com/articles/20174631}).

	\item {\bf Newsletter and RSS feeds}, which refer to e-mails with the most relevant news about the projects and RSS feeds.

\end{itemize}

In addition, we found that developers use Q\&A forums (e.g., StackOverflow), discussion groups (e.g., Google Groups), and messaging tools (e.g., IRC and Slack) to promote their projects.
However, these channels are mostly used to discuss the projects and to provide answers to common questions raised by users.
For example, from the 155 topics opened in 2017 in the {\sc adobe/brackets} discussion group at Google Groups, only eight (5.1\%) are related to announcements of new versions, mostly pre-releases for community testing.
Moreover, from almost 500 topics on {\sc facebook/react} official forum, we could not identify any announcement related to the project development.
Thus, in this study, we do not consider forums, discussion groups, and messaging tools as promotion channels.

\section{Results}
\label{results}

\subsection{What are the most common promotion channels?}
\label{sec:results:rq1}

\begin{figure}[!ht]
	\centering
	\includegraphics[width=0.75\textwidth,keepaspectratio,trim={0 1em 0 3em},clip]{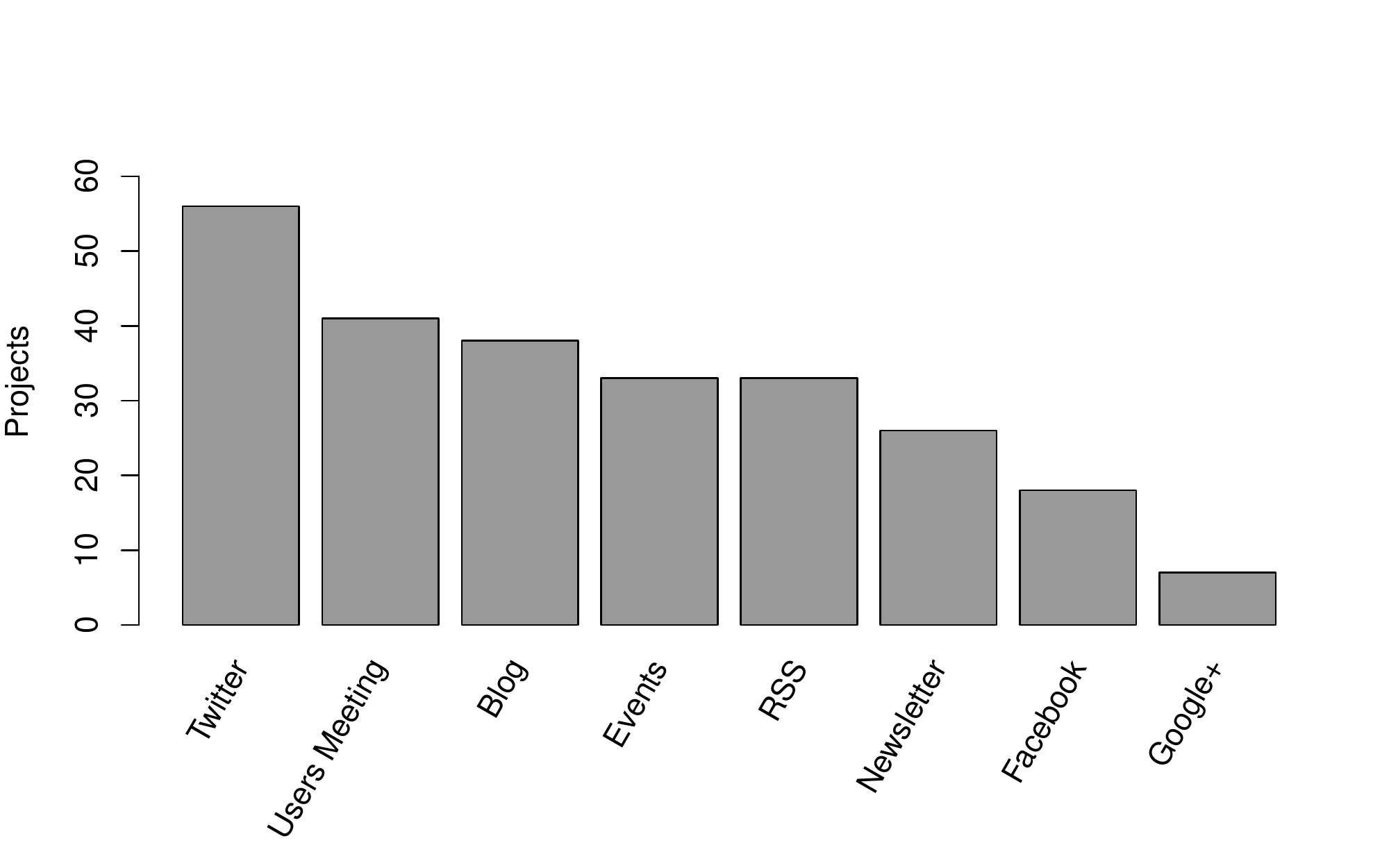}
	\caption{Most common promotion channels}
	\label{fig:rq1}
\end{figure}

Figure~\ref{fig:rq1} presents the most common promotion channels used by the top-100 projects on GitHub.
The most common channel is Twitter, which is used by 56 projects.
The second one is Users Meetings (41 projects), followed by Blogs (38 projects), Events (33 projects), and RSS feeds (33 projects).
The least common channels are Facebook and Google+, which are used by 18 and 7 projects, respectively.

\begin{figure}[!ht]
	\centering
	\includegraphics[width=0.725\textwidth,keepaspectratio,trim={0 1em 0 3em},clip]{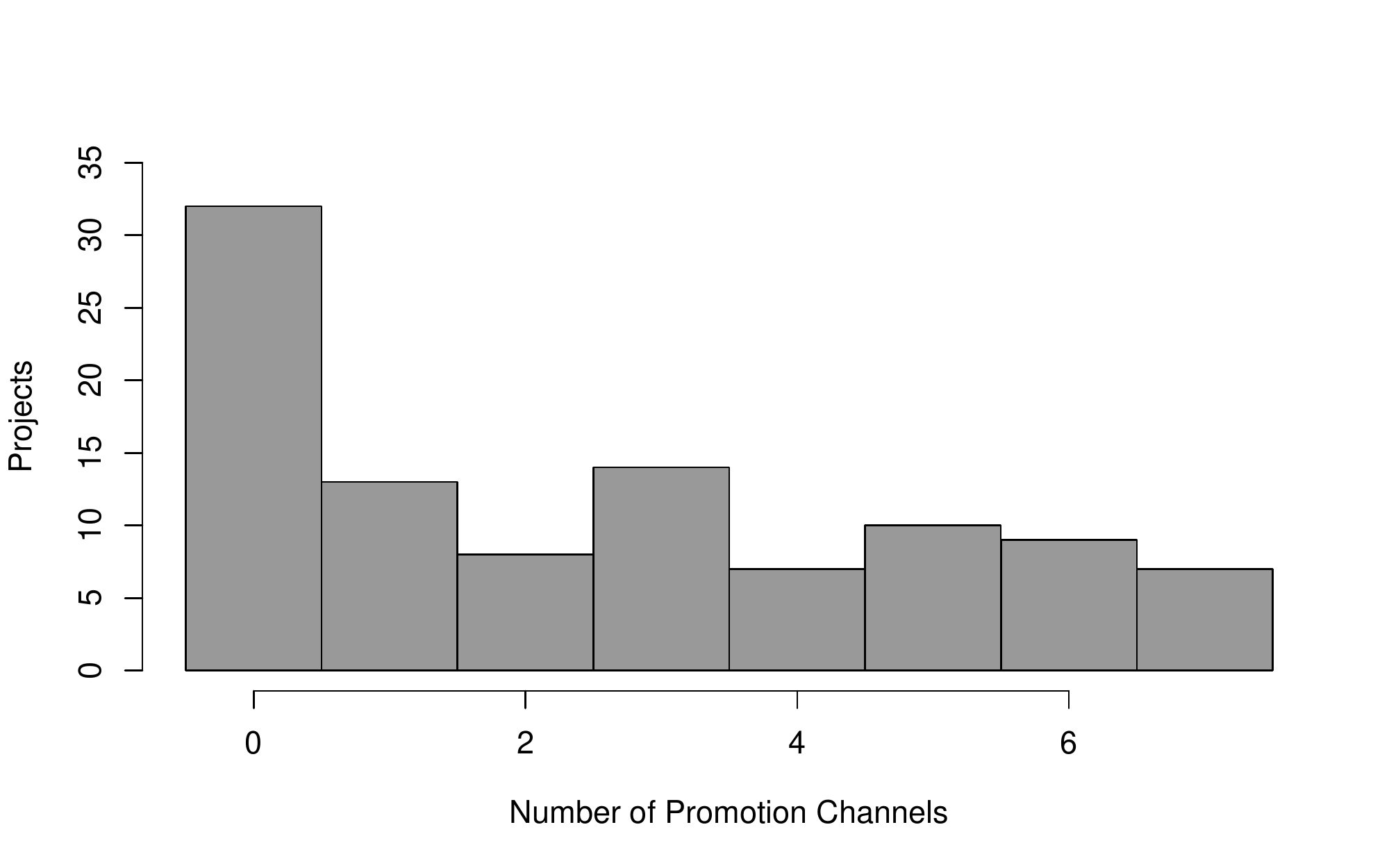}
	\caption{Number of promotion channels per project}
	\label{fig:rq1_2}
\end{figure}

Figure~\ref{fig:rq1_2} shows the distribution of the number of promotion channels per project.
Almost one third of the projects (32 projects) do not use any channel.
By contrast, more than half of the projects (55 projects) use at least two promotion channels.
The highest number of promotion channels is seven, which is the case of {\sc \mbox{facebook/react}}, {\sc \mbox{facebook/react-native}}, {\sc \mbox{meteor/meteor}}, {\sc \mbox{golang/go}}, {\sc \mbox{ionic-team/ionic}}, {\sc \mbox{angular/angular}}, and {\sc adobe/\\brackets}.
We also found that Blog and Twitter is the most frequent combination of channels (35 projects).
Other frequent combinations include, for example, Blog and RSS (31 projects), Events and Users Meetings (31 projects), and Twitter, Events and User Meetings (31 projects).

\subsection{How often do developers promote their projects?}
\label{sec:results:rq2}

In this second question, we investigate how often developers promote their projects on blogs and social networks.
For blogs, we calculate the promotion frequency as the number of posts on the last 12 months.
For social networks, we could not retrieve all posts for all projects because their  APIs restrict the search to a recent period (e.g., last seven days for Twitter and last 100 posts for Facebook).
Thus, in this case, we only classified each social network account in two distinct groups: active and inactive.
An {\em active} account has at least three posts on the last three months; otherwise, it is considered an {\em inactive} account.
This classification was performed by manually counting the number of posts on the social network pages.

\begin{figure}[!ht]
	\centering
	\includegraphics[width=.325\linewidth,keepaspectratio,trim={0 2em 0 2em},clip,page=2]{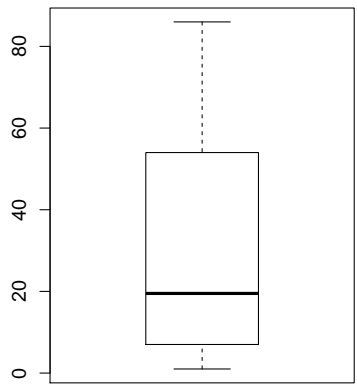}
	\caption{Distribution of the number of posts on the last 12 months (outliers are omitted)}
	\label{fig:rq2}
\end{figure}

Figure~\ref{fig:rq2} presents the distribution of the number of blog posts on the last 12 months.
The number ranges from 1 ({\sc nylas/nylas-mail}) to 1,300 ({\sc freeCodeCamp/freeCodeCamp}); the first, second, and third quartile values are 7, 19, and 54 posts, respectively.

Table~\ref{tab:rq2:social} lists the activity status of the Twitter, Facebook, and Google+ accounts.
We found that 83.9\% of the projects that use Twitter have an active account; 55.6\% of the projects have an active Facebook account and only 28.6\% have an active Google+ account.

\begin{table}[!ht]
  \caption{Active Twitter, Facebook, and Google+ accounts}
  \label{tab:rq2:social}
  \centering
  \begin{tabular}{@{}ccrr@{}}
    \toprule
    \multicolumn{1}{c}{\bf Channel} && \multicolumn{1}{c}{\bf Active (\%)} & \multicolumn{1}{c}{\bf Inactive (\%)} \\
    \midrule
     Twitter  && 47 (83.9\%) & 9 (16.1\%) \\
     Facebook && 10 (55.6\%) & 8 (44.4\%) \\
     Google+  && 2  (28.6\%) & 5 (71.4\%) \\
    \bottomrule
  \end{tabular}
\end{table}

\begin{figure}[!ht]
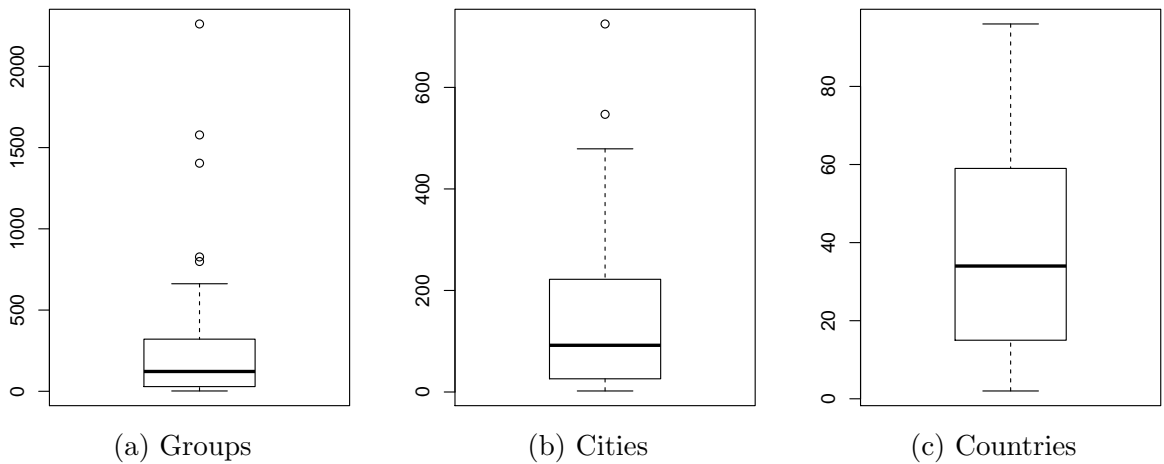

	\centering
	\begin{subfigure}{.325\textwidth}
	  \centering
	  \includegraphics[width=.95\linewidth,keepaspectratio,trim={1.5em 2em 1em 2em},clip,page=2]{images/meetups.pdf}
		\caption{Groups}
	  \label{fig:rq3_meetups_sub1}
	\end{subfigure}%
	\begin{subfigure}{.325\textwidth}
	  \centering
	  \includegraphics[width=.95\linewidth,keepaspectratio,trim={1.5em 2em 1em 2em},clip,page=3]{images/meetups.pdf}
		\caption{Cities}
	  \label{fig:rq3_meetups_sub2}
	\end{subfigure}%
	\begin{subfigure}{.325\textwidth}
	  \centering
	  \includegraphics[width=.95\linewidth,keepaspectratio,trim={1.5em 2em 1em 2em},clip,page=4]{images/meetups.pdf}
		\caption{Countries}
	  \label{fig:rq3_meetups_sub3}
	\end{subfigure}%
	\caption{Number of groups, cities, and countries of the user meetings}
	\label{fig:rq3_meetups}
\end{figure}

Finally, we investigate the characteristics of the user meeting groups promoted on Meetup (such meetings are the 3rd most common promotion channel studied in this article).
A Meetup group is a local community of people that is responsible for organizing meeting events~\cite{meetupgroup}. 
These groups are identified by topics to help members find them. 
Here, we rely on these topics to collect meetups about the studied open source projects, along with their locations (i.e., city and country).
For example, the topic for {\sc jquery/jquery}  is {\em jquery} and a summary of the meeting groups about this topic can be found at \url{https://www.meetup.com/topics/jquery/all}.
Figure~\ref{fig:rq3_meetups} presents the distribution of the number of groups, cities, and countries of the projects with meetings registered at Meetup. For groups, the values ranges from 2 to 2,261 groups; considering the cities, the values range from 2 to 725; finally, for countries, the values range from 2 to 96. The maximum values always refer to {\sc torvalds/linux}. In other words, {\sc torvalds/linux} has 2,261 meetup groups, which are spread over 725 cities from 96 countries.

\subsection{How popular and random projects differ on the usage of promotion channels?}
\label{sec:results:rq3}

In Section~\ref{sec:results:rq1}, we investigated the most common promotion channels used by popular GitHub projects.
In this section, we contrast the usage of promotion channels by these projects and by a random sample of GitHub projects.
For this purpose, we randomly selected 100 projects from the top-5,000 repositories by number of stars and manually inspected their documentation using the same methodology reported in Section~\ref{sec:design}.
The number of stars of this random sample ranges from 2,297 stars ({\sc uber-archive/image-diff}) to 22,558 ({\sc vsouza/awesome-ios}). 

\begin{figure}[!ht]
	\centering
	\includegraphics[width=0.75\textwidth,keepaspectratio,trim={0 1em 0 2em},clip]{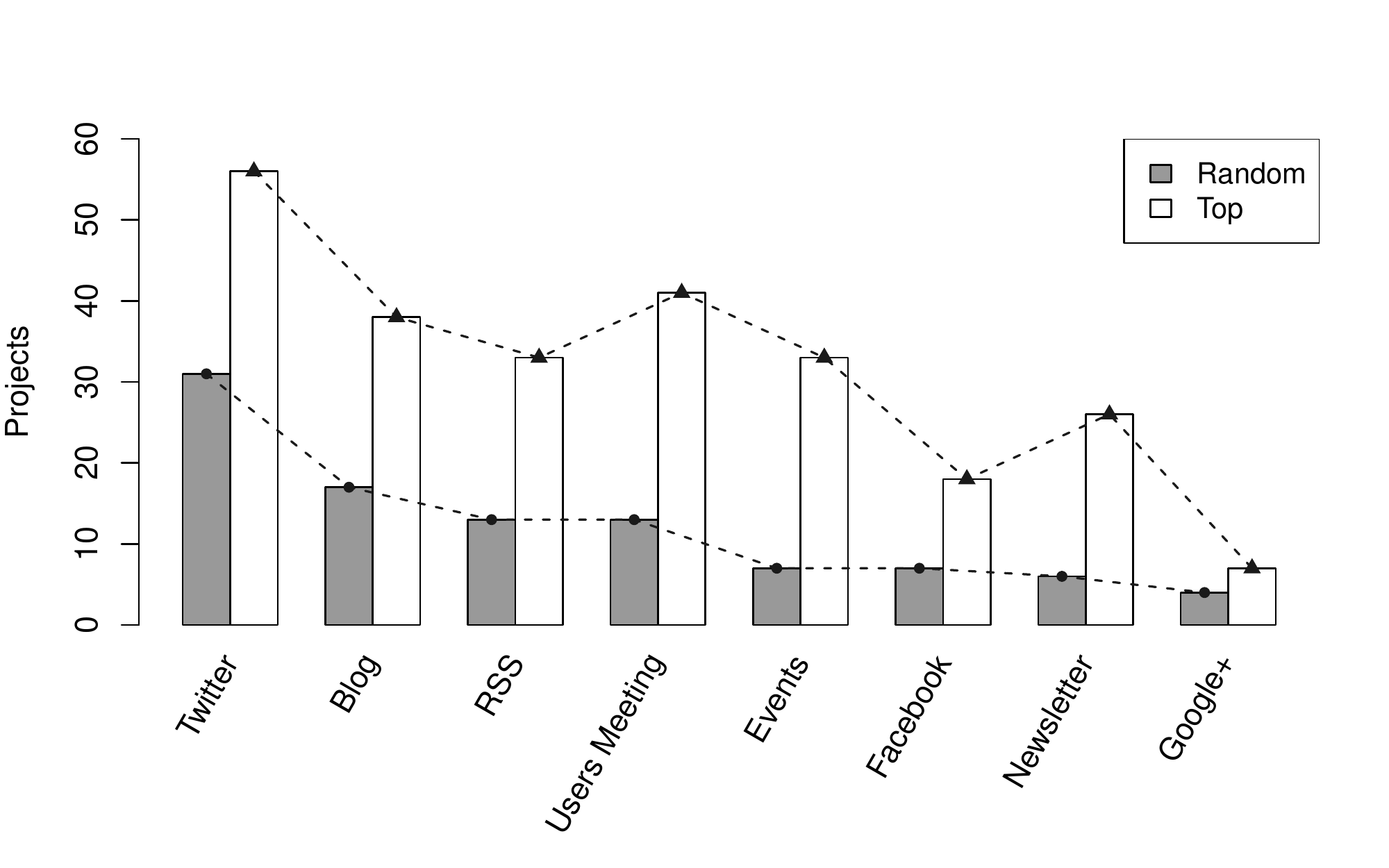}
	\caption{Most common promotion channels used by random projects}
	\label{fig:rq3}
\end{figure}

Figure~\ref{fig:rq3} compares the usage of promotion channels by the random projects and by the most popular ones.
In the random sample, the number of projects using the investigated promotion channels is significantly lower compared to the most popular ones.
However, by applying the Spearman's rank correlation test, we found a strong correlation between the number of projects using the promotion channels on each group ($rho =$ 0.904 and \emph{p-value} $<$ 0.01).
For example, Twitter is also the most used promotion channel among the random projects (31 projects), followed by Blogs (17 projects) and RSS (13 projects).
Compared to the most popular projects, Users meetings and Newsletter are less common (13 and 6 projects, respectively).
Finally, Facebook and Google+ also have a very limited usage (7 and 4 projects, respectively).

\subsection{What is the impact of promotion on Hacker News?}
\label{results:rq4}

After publishing content on blogs, Twitter, etc., open source developers can also promote this content on social news aggregator sites. These sites aggregate contents from distinct sources for easing viewing by a large public.
The most popular and important example is Hacker News (\url{https://news.ycombinator.com}), which is dedicated to Computer Science and related technologies content. Hacker News posts just include a title and the URL of the promoted content (e.g.,~a blog post about a new version of an open source project). Any user registered in the site can post a link on Hacker News, i.e., not necessarily the links are posted by the contributors of an open source project, for example. Other Hacker News users can discuss the posts and upvote
them. An upvote is similar to a {\em like} in social networks; posts are listed on Hacker News according to the number of upvotes.
In this research question, we use Hacker News due to its popularity;  posts that reach the front page of the site receive for example 10-100K page views, in one or two days (\url{https://goo.gl/evyP4w}). Furthermore, Hacker News
provides a public API, which allows search and metadata collection.

\begin{figure}[!ht]
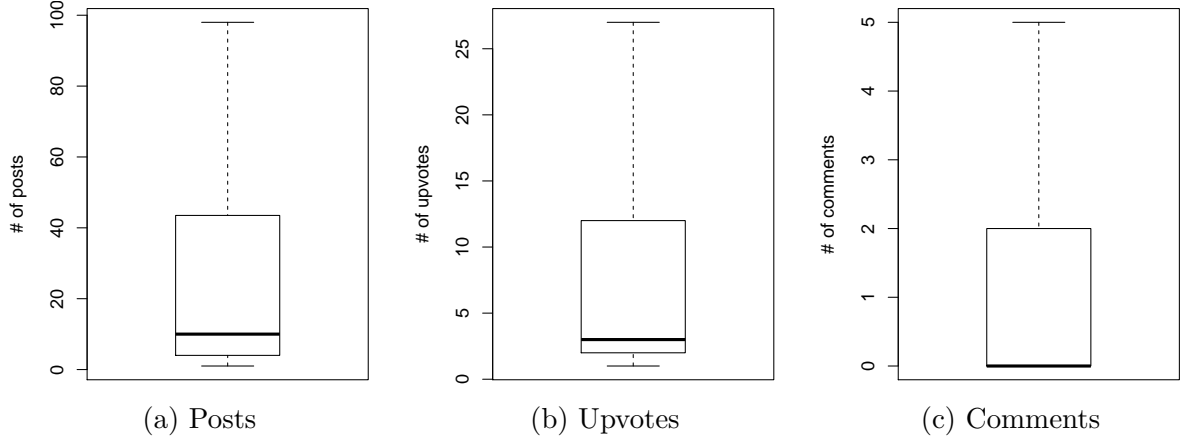

	\centering
	\begin{subfigure}{.325\textwidth}
	  \centering
	  \includegraphics[width=.95\linewidth,keepaspectratio,trim={0 2em 1em 2em},clip,page=2]{images/hn_posts_overview.pdf}
		\caption{Posts}
	  \label{fig:rq4_overview_sub1}
	\end{subfigure}%
	\begin{subfigure}{.325\textwidth}
	  \centering
	  \includegraphics[width=.95\linewidth,keepaspectratio,trim={0 2em 1em 2em},clip,page=4]{images/hn_posts_overview.pdf}
		\caption{Upvotes}
	  \label{fig:rq4_overview_sub2}
	\end{subfigure}%
	\begin{subfigure}{.325\textwidth}
	  \centering
	  \includegraphics[width=.95\linewidth,keepaspectratio,trim={0 2em 1em 2em},clip,page=6]{images/hn_posts_overview.pdf}
		\caption{Comments}
	  \label{fig:rq4_overview_sub3}
	\end{subfigure}%
	\caption{Number of posts, upvotes, and comments (outliers are omitted)}
	\label{fig:rq4_overview}
\end{figure}

For each popular project considered in our study (100 projects), we searched for Hacker News posts with a URL referencing the project sites or pages, including GitHub pages (READMEs, issues, etc). As result, we found 3,019 posts on Hacker News referencing content from 96 studied projects (i.e., only four projects are never referenced on Hacker News).
Figure~\ref{fig:rq4_overview} presents the distributions of the number of posts per project, upvotes, and comments.
The number of posts ranges from 1 to 298 posts per project ({\sc rails/rails}); the first, second, and third quartile values are 4, 10, and 43 posts, respectively.
Regarding their upvotes, the most popular post is about {\sc appple/swift} (\aspas{\em Swift is Open Source}), with 1,824 upvotes; the quartile values are 2, 3, and 12 upvotes, respectively.
Finally, the highest number of comments is 760, about a GitHub issue opened for Microsoft Visual Studio (\aspas{\em VS Code uses 13\% CPU when idle due to blinking cursor rendering}); the quartile values are 0, 0, and 2 comments, respectively. 
On the one hand, these results show that most Hacker News posts do not attract attention. 
By contrast, a small number of posts attract a lot of attention. For example, the top-10\% posts have at least 132 upvotes. These posts are called {\em successful posts} in this investigation.

\begin{figure}[!ht]
	\centering
	\includegraphics[width=0.4\textwidth,keepaspectratio,trim={0 0 0 2em},clip,page=2]{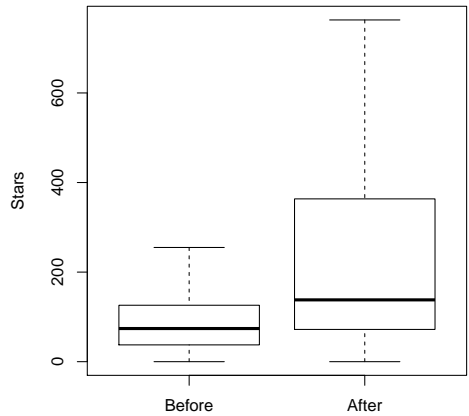}
	\caption{Number of GitHub stars received by projects covered by successful Hacker News posts in the first three days before and after the post publication}
	\label{fig:rq4_stars_before_after}
\end{figure}

Figure~\ref{fig:rq4_stars_before_after} shows boxplots with the number of GitHub stars gained by projects covered by successful posts, in the first three days before and  after the publication date on Hacker News. The  intention is to investigate the impact of a successful promotion on Hacker News, by comparing the number of stars gained before and after each successful post publication. On the median, the projects covered by successful posts gained 74 stars in the first three days before their appearance on Hacker News; in the first three days after the publication, the projects gained 138 stars. Therefore, Hacker News has a positive impact on the project's popularity, measured by GitHub stars. 
Indeed, the distributions are statistically different, according to the one-tailed variant of the Mann-Whitney U test (p-value $\leq 0.05$). By computing Cliff's delta, we found a {\em medium} effect size ($d = -0.372$).

Finally, we inspected the titles of each successful post, aiming to categorize the post purpose. The most common category includes posts announcing new releases of open source projects (44.9\%; e.g., \aspas{\em Angular 2 Final Released}). Other popular categories include posts promoting articles or reports about the projects (25.4\%; e.g., \aspas{\em Vue.js vs.~React}), announcing the first release of a project (16.5\%; e.g., \aspas{\em YouTube-dl: Open-source YouTube downloader}), highlighting new project features (10.6\%; e.g., \aspas{\em Git and GitHub Integration Comes to Atom}) and open sourcing products (1.6\%; e.g., \aspas{\em Visual Studio Code is now open source}).

\section{Related Work}
\label{sec:related}

Although open source software has been exhaustively explored recently, little is known about how developers promote these projects.
The main exception is a work conducted by Bianco et al. where the authors analyze marketing and communication strategies of three companies that develop open source software~\cite{Bianco12}.
By means of interviews, they found that websites and product launch events are adopted by the three organizations; however, the organizations differ considerably on the use of other communication channels, mainly when promoting the projects in open source communities and among industrial users.

On the one hand, most communication channels investigated in this paper are explored in other studies, but with different intentions.
Singer et al. report a qualitative study focused on discovering the benefits that Twitter brings to developers~\cite{singer2014}.
They found that Twitter adopters use it to stay aware of industry changes, for learning, and for building relationships.
By correlating the blogging and committing behavior of developers, Pagano and Maleej observed an intensive use of blogs, frequently detailing activities described shortly before in commit messages~\cite{pagano2011}.
Bajic and Lyons analyze how software companies use social media techniques to gather feedback from users collectively~\cite{Bajic2011}.
Their results suggest that startups use social media mainly for competitive advantage and established organizations use it to monitor the buzz among their users.
By studying a successful software development company, Hansson et al. identified that user meetings and newsletter are adopted to include and increase the participation of users in the development process~\cite{Hansson2006}.
Finally, Aniche et al. conduct a study to understand how developers use modern news aggregator sites (Reddit and Hacker News)~\cite{aniche2018}. According to their results, the two main reasons for posting links on these sites is to promote own work and to share relevant content.

\section{Conclusion and Practical Implications}
\label{sec:conclusion}

In this paper, we investigated the most common promotion channels used by popular GitHub projects. This investigation supports the following practical recommendations to open source project managers and leaders:

\begin{enumerate}
\item Promotion is an important aspect of open source project management, which should be emphasized by project leaders. For example, most popular GitHub projects (two thirds) use at least one promotion channel; half of the projects invest on two channels. By contrast, the use of promotion channels is less common among projects with lower popularity. 

\item Open source project managers should  consider  the use of Twitter (47 projects among the top-100 most popular GitHub projects have active Twitter accounts), Users meetings (which are organized or supported by 41 projects), and blogs (which are used by 38 projects).

\item Open source project managers should also consider promotion on social news aggregator sites. Successful posts on Hacker News may have an important impact on the popularity of GitHub projects. However, only 10\% of the Hacker News posts about the studied projects have had some success
\end{enumerate}


\section*{Acknowledgments}

\noindent Our research is supported by CAPES, FAPEMIG, and CNPq.

\small
\bibliographystyle{IEEEtran}
\bibliography{bibfile}

\end{document}